\documentclass[american,aps,pra,reprint,floatfix,nofootinbib,superscriptaddress]{revtex4-1}
\usepackage[unicode=true,pdfusetitle, bookmarks=true,bookmarksnumbered=false,bookmarksopen=false, breaklinks=false,pdfborder={0 0 0},backref=false,colorlinks=false]{hyperref}
\hypersetup{colorlinks,linkcolor=myurlcolor,citecolor=myurlcolor,urlcolor=myurlcolor}
\usepackage{graphics,epstopdf,graphicx,amsthm,amsmath,amssymb,mathptmx,braket,colortbl,color,bm,framed,mathrsfs}
\usepackage{graphicx}
\usepackage[T1]{fontenc}
\usepackage[up]{subfigure}
\usepackage{tikz}
\usetikzlibrary{decorations.pathmorphing}
\definecolor{myurlcolor}{rgb}{0,0,0.9}

\newcommand{\proj}[1]{| #1\rangle\!\langle #1 |}

\newcommand{\iinner}[2]{\langle #1 | #2\rangle}

\DeclareMathOperator{\trace}{Tr}
\newcommand{\Ptr}[2]{\trace_{#1}\Pa{#2}}
\newcommand{\Tr}[1]{\Ptr{}{#1}}
\newcommand{\Innerm}[3]{\left\langle #1 \left| #2 \right| #3 \right\rangle}
\newcommand{\Pa}[1]{\left[#1\right]}

\newcommand{\norm}[1]{\left\lVert #1 \right\rVert}
\newcommand{\pa}[1]{\left(#1\right)}

\theoremstyle{plain}

\newcommand*{\myproofname}{Proof}

\def\ot{\otimes}
\def\complex{\mathbb{C}}

\def\cM{\mathcal{M}}

\def\md{{(\mathrm{d})}}

\makeatother

\begin{document}

  \author{Kaifeng Bu}
 \email{bkf@zju.edn.cn}
 \affiliation{School of Mathematical Sciences, Zhejiang University, Hangzhou 310027, PR~China}
  \author{Lu Li}
 \affiliation{School of Mathematical Sciences, Zhejiang University, Hangzhou 310027, PR~China}

 \author{Junde Wu}
 \affiliation{School of Mathematical Sciences, Zhejiang University, Hangzhou 310027, PR~China}

   \author{Shao-Ming Fei}
    \email{feishm@cnu.edn.cn}
 \affiliation{School of Mathematical Sciences, Capital Normal University, Beijing 100048, PR China}
  \affiliation{Max-Planck-Institute for Mathematics in the Sciences, 04103 Leipzig, Germany}

\title{Duality relation between coherence and path information in the presence of quantum memory }

\begin{abstract}
The wave-particle duality demonstrates a competition relation between wave and particle behavior for
a particle going through an interferometer. This duality can be formulated as an inequality, which  upper
bounds the sum of interference visibility and path information. However, if the particle is
entangled with a quantum memory, then the bound may decrease. Here, we find the duality relation between  coherence
and path information for a particle going through a multipath interferometer in the presence of a quantum memory, offering an upper bound on the duality relation which is directly connected with the
amount of entanglement between the particle and the quantum memory.
\end{abstract}

\maketitle

\section{Introduction}
Quantum coherence, defined as the degree of superposition in a given reference basis,  can be
used to characterize the quantumness in a single system, and plays an important role
in a variety of applications, ranging from  metrology \cite{Giovannetti2011} to thermodynamics \cite{Lostaglio2015,Lostaglio2015NC}.
Recently, the development of a resource theory of coherence has attracted much attention \cite{Baumgratz2014,Girolami2014,Streltsov2015,Winter2016,Killoran2016,Chitambar2016,Chitambar2016a}.
 One of the main advantages that a resource theory offers is the lucid quantitative and operational description at ones disposal.
In order to quantify the amount of coherence, two coherence measures have been proposed, namely,  $l_1$ norm of coherence and
relative entropy of coherence \cite{Baumgratz2014}.  As coherence can be used to characterize the wave behavior of a particle,
here we investigate the duality relation between coherence and path information for a particle  going through a multipath interferometer  with
the access to a quantum memory, where the coherence is quantified by $l_1$ norm and relative entropy of coherence.

The wave-particle duality illustrates that a particle can exhibit both wave and particle behavior when it goes through an interferometer. A number of
quantitative formulations for this duality  have been proposed \cite{Wootters1979,Greenberger1988,Jaeger1995,Englert1996, Englert2000,Englert2008,Liu2009,Li2012,Huang2013,Banaszek2013}. One well-known duality relation for a two-path
interferometer is given in \cite{Jaeger1995,Englert1996} as follows,
\begin{eqnarray}\label{eq:WPdual}
D^2+V^2\leq 1,
\end{eqnarray}
where particle behavior is quantified by the path information (or path distinguishability) $D$  and wave behavior is
quantified by the interference visibility $V$. This tradeoff relation shows that the path information $D$ will give a limitation
on the interference visibility $V$ and vice versa. Besides, the connection between wave-particle duality and Heisenberg' uncertainty principle has been investigated \cite{Storey1994,Englert1995,Wiseman1995,Coles2014},  where Heisenberg' uncertainty principle demonstrates that the complementary observables cannot
be measured precisely at the same time.
The equivalence between these two concepts in certain formulation, where the particle and wave behavior are captured by the measurements on the  so-called particle and wave  observables, illustrates the significance of wave-particle duality in the operational tasks \cite{Coles2014}.

The  wave-particle duality for multipath interferometers  is first investigated in \cite{Durr2001} in terms of the density matrix of the particle represented in the
path basis. In a similar scenario, the duality relations between path coherence and path information have  been proposed in terms of the coherence measures in the resource theory
of coherence \cite{Bera2015,Bagan2016,Paul2017}. For a given reference basis $\set{\ket{i}}^{N}_{i=1}$, $l_1$ norm of coherence  is defined as $C_{l_1}(\rho)=\sum_{i\neq j}|\rho_{ij}|$ with $\rho_{ij}=\Innerm{i}{\rho}{j}$ and relative entropy of coherence is defined as
 $C_r(\rho)=S(\rho^\md)-S(\rho)$ where $S(\rho)=-\Tr{\rho\log\rho}$ is the von Neumann entropy  and $\rho^\md=\sum_i\rho_{ii}\proj{i}$ is the diagonal part of $\rho$ \cite{Baumgratz2014}. The wave behavior of the particle is quantified by these two coherence measures.
  To detect path information, detectors are used to interact with the particle and the path information is quantified by the discrimination
of detector states \cite{Bera2015,Bagan2016}. It has been proved that the sum of path coherence and the optimal success probability of discriminating detector states (by unambiguous state discrimination)  is less than or equal to one \cite{Bera2015}. Later an improved duality relation between coherence and path information in N-path
interferometer is proved, with the path information quantified by the ambiguous (or minimal error) state discrimination,
\begin{eqnarray}\label{eq:CPdual}
(P_s-\frac{1}{N})^2+X^2\leq (1-\frac{1}{N})^2,
\end{eqnarray}
where $P_s$ is the optimal success probability to discriminate detector states by ambiguous state discrimination and $X$ is the normalized $l_1$ norm coherence of the particle defined by $X=\frac{1}{N}C_{l_1}(\rho)$ \cite{Bagan2016}. The quantity
$P_s-1/N$ quantifies the advantage to discriminate detector states by prior knowledge compared with random guessing \cite{Bagan2016}.

\begin{figure}
\centering
     \begin{tikzpicture}[scale=0.2]
    \begin{scope}[thick,color=black] 
        \draw (0,0)--(0,5);
       \draw (0,6)--(0,9);
       \draw (0,10)--(0,13);
       \draw (0,14)--(0,17);
       \draw (0,18)--(0,23) node[above] {N-path};
    \end{scope}
 \begin{scope}[color=black] 
        \draw (12,0)--(12,23)node[above] {Screen};
         \draw[densely dashed](13,1.5)
    	         to [out=90,in=270] (14,3.5);
        \draw[densely dashed](14,3.5)
    	         to [out=90,in=270] (13,5.5);
         \draw[densely dashed](13,5.5)
    	         to [out=90,in=270] (15,7.5);
        \draw[densely dashed](15,7.5)
    	         to [out=90,in=270] (13,9.5);
      \draw[densely dashed](13,9.5)
    	         to [out=90,in=270] (16.5,11.5);
        \draw[densely dashed](16.5,11.5)
    	         to [out=90,in=270] (13,13.5);
      \draw[densely dashed](13,13.5)
    	         to [out=90,in=270] (15,15.5);
        \draw[densely dashed](15,15.5)
    	         to [out=90,in=270] (13,17.5);
       \draw[densely dashed](13,17.5)
    	         to [out=90,in=270] (14,19.5);
    \draw[densely dashed](14,19.5)
    	         to [out=90,in=270] (13,21.5);
    \end{scope}
    \begin{scope}[color=red] 
        \draw (1,0) rectangle node {D}
            ++(1.3,23);
         \draw (2,-0.2)node[below]{Detector};
    \end{scope}
   \begin{scope}[color=blue] 
          \draw (-6,-9) rectangle node {Memory B}
            ++(-8,4);
         \draw[densely dashed,gray,font=\small,->] (-10,11.5) -- (0,5.5);
         \draw[densely dashed,gray,font=\small,->] (-10,11.5) -- (0,9.5);
         \draw[densely dashed,gray,font=\small,->] (-10,11.5) -- (0,13.5);
         \draw[densely dashed,gray,font=\small,->] (-10,11.5) -- (0,17.5);
          \draw[thick,decorate, decoration=snake](-10,-5)
    	         to(-10,11.5);
     \filldraw (-10,11.5) circle (0.5) node[left=5] {Particle A};
    \end{scope}
    \end{tikzpicture}
  \caption{Schematic diagram of an N-path interferometer. The particle A goes through an N-path interferometer while A is entangled with a quantum memory B.
  The detector D is used to detect which path the particle A goes through.}
  \label{fig1}
\end{figure}
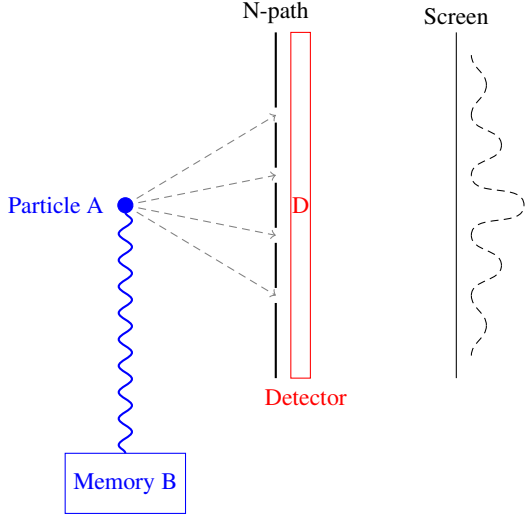

Equation \eqref{eq:CPdual} bounds the duality between  coherence and path information without quantum memory. However, this bound may
decrease if the particle is entangled with a quantum memory. Here, we show a duality relation between coherence and path information for a particle A going through
an N-path interferometer in the presence of a quantum memory B (see Fig.\ref{fig1}), which provides an upper bound on the sum of the coherence and path information
that depends on the amount of entanglement between the particle A and the quantum memory B,
\begin{equation}\label{eq:ECPdual}
\begin{array}{rcl}
\pa{P^A_s-\frac{1}{N}}^2+X^2_A&\leq& \pa{1-\frac{1}{N}}^2\\[2mm]
&&+\frac{2(N-1)}{N^2}\pa{\Tr{\rho^2_A}-\Tr{\rho^2_{AB}}}.
\end{array}
\end{equation}
The extra term $\Tr{\rho^2_A}-\Tr{\rho^2_{AB}}$ on the right-hand side of \eqref{eq:ECPdual} quantifies the amount of entanglement between particle A and the memory B, as
\begin{eqnarray}\label{ineq:ent}
\Tr{\rho^2_A}-\Tr{\rho^2_{AB}}<0
\end{eqnarray}
only if $\rho_{AB}$ is entangled \cite{Horodecki1996}. Inequality \eqref{ineq:ent}
provides a powerful tool in experimental detection of entanglement \cite{Bovino2005,Islam2015}. Moreover, for $N=2$, the relation \eqref{eq:ECPdual}
becomes an equality.

Let us begin with a pure bipartite state $\ket{\psi}_{AB}$ between particle A and quantum memory B. Note that we can always consider a pure bipartite state between particle A and a lager quantum memory B$^\prime$ even if the initial bipartite state between A and B is not pure.

In an N-path interferometer, if the orthonormal basis states $\set{\ket{i}}^N_{i=1}$ correspond to the N possible slits or paths, then the state
of the particle A after crossing the slit can be described in terms of $\set{\ket{i}}^N_{i=1}$, and thus
the bipartite pure state $\ket{\psi}_{AB}$ can be represented as
$\ket{\psi}_{AB}=\sum^{N}_{i=1}\sum^{d_B}_{j=1}a_{ij}\ket{i}_A\ket{j}_B$, with $a_{ij}\in\complex$, $\sum_{i,j}|a_{ij}|^2=1$ and $d_B$ being the dimension of quantum memory B.
Denote $p_i=\sum_{j}|a_{ij}|^2$ and $\ket{u_i}_B=\sum_j a_{ij}\ket{j}_B/\sqrt{p_i}$. $\ket{\psi}_{AB}$ can also be written as
\begin{eqnarray}\label{eq:iniS}
\ket{\psi}_{AB}=\sum^{N}_{i=1}\sqrt{p_i}\ket{i}_A\ket{u_i}_B,
\end{eqnarray}
where $p_i$ is the probability to take the $i$th path and $\ket{u_i}_B$ is the normalized
pure state on memory B for any $i\in\set{1,...,N}$. Here $\ket{u_i}_B$ are not necessary orthogonal.
In order to detect particle A, another quantum system called detector  interacts with the particle A inside the interferometer, and the interaction is
described as the controlled unitary $U(\ket{i}_A\ket{\phi_0}_D)=\ket{i}_A\ket{\phi_i}_D$ with $\ket{\phi_0}_D$ being the initial state
of the detector. After the interaction, the state of the  whole system becomes
\begin{eqnarray}\label{den_w}
\ket{\Psi}_{ABD}=\sum^N_{i=1}\sqrt{p_i}\ket{i}_A\ket{u_i}_B\ket{\phi_i}_D.
\end{eqnarray}
Thus the reduced density matrix of the combined system $A,B$ is
\begin{eqnarray}\label{den_AB}
\nonumber\rho_{AB}&=&\Ptr{D}{\proj{\Psi}_{ABD}}\\
&=&\sum^{N}_{i,j=1}\sqrt{p_ip_j}\iinner{\phi_j}{\phi_i}\ket{i}\bra{j}_A\ot\ket{u_i}\bra{u_j}_B,
\end{eqnarray}
and the reduced density matrix of the particle A is
\begin{eqnarray}\label{den_A}
\nonumber\rho_A&=&\Ptr{BD}{\proj{\Psi}_{ABD}}\\
&=&\sum^{N}_{i,j=1}\sqrt{p_ip_j}\iinner{\phi_j}{\phi_i}\iinner{u_j}{u_i}\ket{i}\bra{j}_A.
\end{eqnarray}
Hence the normalized $l_1$ norm of  coherence measure $X_A$ is
\begin{eqnarray}
X_A=\frac{1}{N}C_{l_1}(\rho_A)
=\frac{1}{N}\sum^{N}_{\substack{
i,j=1\\
i\neq j}
}\sqrt{p_ip_j}|\iinner{\phi_j}{\phi_i}||\iinner{u_j}{u_i}|.
\end{eqnarray}
Besides, the path information is stored in the detector states and the density matrix of the detector can be expressed as
\begin{eqnarray}\label{den_D}
\rho_D=\Ptr{AB}{\proj{\Psi}_{ABD}}
=\sum^N_{i=1}p_i\proj{\phi_i}_D.
\end{eqnarray}
In order to obtain the path information, we need perform state discrimination on the set of quantum states $\set{\ket{\phi_i}_D}^N_{i=1}$ with
the probability $\set{p_i}^N_{i=1}$. For a positive operator valued measure (POVM)  $\set{\Pi_i}^N_{i=1}$ with $\Pi_i\geq 0$ and $\sum_i\Pi_i=\mathbb{I}$, the success probability  to discriminate the states $\set{\ket{\phi_i}_D}$  is $\sum^N_{i=1}p_i\Tr{\Pi_i\proj{\phi_i}}$. Thus the optimal success  probability among all POVMs is
\begin{eqnarray}
P^A_s=\max_{\set{\Pi_i}}\sum^N_{i=1}p_i\Tr{\Pi_i\proj{\phi_i}}.
\end{eqnarray}
For ambiguous state discrimination, an upper bound for the optimal
success probability  $P^A_s$ is given by \cite{Bagan2016},
\begin{eqnarray}\label{ineq:path1}
P^A_s\leq\frac{1}{N}+\frac{1}{2N}\sum^N_{i,j=1}\norm{T_{ij}}_1,
\end{eqnarray}
where the operators $\set{T_{ij}}$ are given by $T_{ij}=p_i\proj{\phi_i}-p_j\proj{\phi_j}$ and trace norm of $T_{ij}$ for $i\neq j$ is given by
\begin{eqnarray*}
\norm{T_{ij}}_1=2\sqrt{\pa{\frac{p_i+p_j}{2}}^2-p_ip_j|\iinner{\phi_i}{\phi_j}|^2}.
\end{eqnarray*}
Here, the quantum memory B is not involved  in the discrimination of the detector states, as the
quantum memory may be far away from the interferometer and one may not be able to perform a joint measurement on both the memory B and
the detector D. Note that, if the joint measurement on B and D is allowed, then the memory B would be able to increase the optimal success probability to discriminate states and  the relation \eqref{eq:ECPdual} will reduce to  \eqref{eq:CPdual} following the method in \cite{Bagan2016}.

The bound \eqref{ineq:path1} gives an upper bound for the sum of path information
$\pa{P^A_s-1/N}^2$ and the square of the normalized $l_1$ norm of coherence $X^2_A$
 for particle A,
\begin{eqnarray}\label{ineq:1}
\nonumber\pa{P^A_s-\frac{1}{N}}^2+X^2_A
\leq  \frac{1}{N^2}
\pa{\sum^{N}_{\substack{
i,j=1\\
i\neq j}
}\frac{1}{2}\norm{T_{ij}}_1 }^2\\
+\frac{1}{N^2}\pa{\sum^{N}_{\substack{
i,j=1\\
i\neq j}
}\sqrt{p_ip_j}|\iinner{\phi_i}{\phi_j}||\iinner{u_i}{u_j}| }^2.
\end{eqnarray}
Due to the Schwarz inequality, we can get some upper bounds for the
two terms on the right-hand side of \eqref{ineq:1}, where the trace norm of
$T_{ij}$ is upper bounded as follows,
\begin{eqnarray}\label{ineq:2}
&&\nonumber\pa{\sum^{N}_{\substack{
i,j=1\\
i\neq j}
}\frac{1}{2}\norm{T_{ij}}_1 }^2\\
\nonumber
&=&\pa{\sum^{N}_{\substack{
i,j=1\\
i\neq j}
}\sqrt{\pa{\frac{p_i+p_j}{2}}^2-p_ip_j|\iinner{\phi_i}{\phi_j}|^2} }^2\\\nonumber
&\leq&
\pa{\sum^{N}_{\substack{
i,j=1\\
i\neq j}
}\frac{p_i+p_j}{2}}
\pa{\sum^{N}_{\substack{
i,j=1\\
i\neq j}
}\frac{p_i+p_j}{2}-2\frac{p_ip_j}{p_i+p_j}|\iinner{\phi_i}{\phi_j}|^2 }\\
&=&(N-1)\Pa{(N-1)-\sum^{N}_{\substack{
i,j=1\\
i\neq j}
}2\frac{p_ip_j}{p_i+p_j}|\iinner{\phi_i}{\phi_j}|^2},
\end{eqnarray}
and the $l_1$ norm of coherence is upper bounded as
\begin{eqnarray}\label{ineq:3}
\nonumber&&\pa{\sum^{N}_{\substack{
i,j=1\\
i\neq j}
}\sqrt{p_ip_j}|\iinner{\phi_i}{\phi_j}||\iinner{u_i}{u_j}| }^2\\
\nonumber&\leq& \pa{\sum^{N}_{\substack{
i,j=1\\
i\neq j}
}\frac{p_i+p_j}{2}}
\pa{\sum^{N}_{\substack{
i,j=1\\
i\neq j}
}2\frac{p_ip_j}{p_i+p_j}|\iinner{\phi_i}{\phi_j}|^2|\iinner{u_i}{u_j}|^2}\\
&=& \pa{N-1}
\pa{\sum^{N}_{\substack{
i,j=1\\
i\neq j}
}2\frac{p_ip_j}{p_i+p_j}|\iinner{\phi_i}{\phi_j}|^2|\iinner{u_i}{u_j}|^2},
\end{eqnarray}
where the relation
$\sum^{N}_{i\neq j=1}\frac{p_i+p_j}{2}=N-1$ has been taken into account.
Substituting \eqref{ineq:2} and \eqref{ineq:3} into \eqref{ineq:1}, we have
\begin{eqnarray}\label{ineq:immed1}
\nonumber&&\pa{P^A_s-\frac{1}{N}}^2+X^2_A\\\nonumber
&\leq&
\pa{1-\frac{1}{N}}^2
-2\frac{N-1}{N^2}\sum^{N}_{\substack{
i,j=1\\
i\neq j}
}\frac{p_ip_j}{p_i+p_j}|\iinner{\phi_i}{\phi_j}|^2\pa{1-|\iinner{u_i}{u_j}|^2}\\\nonumber
&\leq& \pa{1-\frac{1}{N}}^2
-2\frac{N-1}{N^2}\sum^{N}_{\substack{
i,j=1\\
i\neq j}
}p_ip_j|\iinner{\phi_i}{\phi_j}|^2\pa{1-|\iinner{u_i}{u_j}|^2}\\
&=&\pa{1-\frac{1}{N}}^2
-2\frac{N-1}{N^2}\pa{\Tr{\rho^2_D}-\Tr{\rho^2_A}},
\end{eqnarray}
where second inequality comes from the fact that
$p_i+p_j\leq 1$ and $|\iinner{u_i}{u_j}|\leq 1$ for any $i\neq j$ and the last equality comes from the fact that
$\Tr{\rho^2_D} $ and $\Tr{\rho^2_A}$ can be expressed as
\begin{eqnarray}\label{eq:MA}
\Tr{\rho^2_A}
=\sum^{N}_{\substack{
i,j=1\\
i\neq j}
}p_ip_j|\iinner{\phi_i}{\phi_j}|^2|\iinner{u_i}{u_j}|^2+\sum^N_{i=1}p^2_i,
\end{eqnarray}
and
\begin{eqnarray}\label{eq:MD}
\Tr{\rho^2_D}
=\sum^{N}_{\substack{
i,j=1\\
i\neq j}
}p_ip_j|\iinner{\phi_i}{\phi_j}|^2+\sum^N_{i=1}p^2_i.
\end{eqnarray}
Besides, since $\rho_{AB}$ and $\rho_D$ are the reduced states of the pure state $\ket{\Psi}_{ABD}$,
the purity of $\rho_{AB}$ and $\rho_{D}$ are equal, $\Tr{\rho^2_{AB}}=\Tr{\rho^2_D}$.
Therefore, we obtain the duality relation  \eqref{eq:ECPdual} in the presence of a quantum memory. In view of the equations \eqref{eq:MA}
and \eqref{eq:MD}, we find that $\Tr{\rho^2_A}\leq \Tr{\rho^2_D}$ which means  $\Tr{\rho^2_A}\leq \Tr{\rho^2_{AB}}$. Thus the
right-hand side of \eqref{eq:ECPdual} is less than or equal to $\pa{1-1/N}^2$. Furthermore, if
$\Tr{\rho^2_A}< \Tr{\rho^2_{AB}}$, then $\rho_{AB}$ is entangled and the right-hand side of \eqref{eq:ECPdual} is strictly less than $\pa{1-1/N}^2$. Note that if the initial bipartite state $\ket{\psi}_{AB}$ is separable,
$\ket{\psi}_{AB}=\ket{\psi'}_A\ket{u}_B$ where $\ket{u_i}_B$ in \eqref{eq:iniS} is equal to $\ket{u}_B$ up to a phase, then $\Tr{\rho^{2}_A}=\Tr{\rho^2_{AB}}$ and the relation \eqref{eq:ECPdual} reduces to \eqref{eq:CPdual}.

For $N=2$, the equality in \eqref{ineq:path1} holds \cite{Helstrom1976}, that is, the optimal success probability
is given by
\begin{eqnarray}\nonumber
P^A_s=\frac{1}{2}+\sqrt{\frac{1}{4}-p_1p_2|\iinner{\phi_1}{\phi_2}|^2},
\end{eqnarray}
and the normalized $l_1$ norm coherence can be written as
\begin{eqnarray}\nonumber
X_A=\sqrt{p_1p_2}|\iinner{\phi_1}{\phi_2}||\iinner{u_1}{u_2}|.
\end{eqnarray}
In this case we have
\begin{eqnarray}
\nonumber\pa{P^A_s-\frac{1}{2}}^2+X^2_A
&=&\frac{1}{4}-p_1p_2|\iinner{\phi_1}{\phi_2}|^2(1-|\iinner{u_1}{u_2}|^2)\\
&=&\frac{1}{4}+\frac{1}{2}\pa{\Tr{\rho^2_{A}}-\Tr{\rho^2_{AB}}}.
\end{eqnarray}
That is, the equality in duality relation \eqref{eq:ECPdual}
holds for two-path interferometer.

If the particle A has no quantum memory and the initial state is $\ket{\tilde{\psi}}_A=\sum^N_{i=1}\sqrt{p_i}\ket{i}_A$, then after the interaction with
the detector,
the reduced state $\tilde{\rho}_A=\sum_{i,j}\sqrt{p_ip_j}\iinner{\phi_j}{\phi_i}\ket{i}\bra{j}_A$ and $\tilde{\rho}_D$ is just given by the equation \eqref{den_D}.
According to \cite{Bagan2016}, the coherence $\tilde{X}_A$ and the path information
$\tilde{P}^A_s$ satisfy the relation \eqref{eq:CPdual}. Compared with the case that A has a quantum memory, the difference between the
bounds of \eqref{eq:CPdual} and \eqref{eq:ECPdual} comes from the loss of coherence of particle in the presence of entanglement, where
the amount of coherence-loss can be quantified as
\begin{equation}
\begin{array}{l}
\frac{1}{N^2}\pa{\Tr{\rho^2_{AB}}-\Tr{\rho^2_A}} \leq \Delta X^2_A\\[3mm]
~~~~~~~~~~~~~~~~~~~~~~~~\leq\frac{2(N-1)^2}{N}\pa{\Tr{\rho^2_{AB}}-\Tr{\rho^2_A}},
\end{array}
\end{equation}
where $\Delta X^2_A=\tilde{X}^2_A-X^2_A$ is the amount of coherence-loss, $\rho_{AB}$ and $\rho_{A}$ are given by \eqref{den_AB} and \eqref{den_A} respectively.
The first inequality comes from the fact that
\begin{eqnarray*}
&&C^2_{l_1}(\tilde{\rho}_A)-C^2_{l_1}(\rho_A)\\
&\geq& \sum^N_{\substack{
i,j=1\\
i\neq j}}p_ip_j|\iinner{\phi_i}{\phi_j}|^2(1-|\iinner{u_i}{u_j}|^2)\\
&=&\Tr{\rho^2_{AB}}-\Tr{\rho^2_A},
\end{eqnarray*}
and the second inequality is due to that
\begin{eqnarray*}
&&C^2_{l_1}(\tilde{\rho}_A)-C^2_{l_1}(\rho_A)\\
&=&(C_{l_1}(\tilde{\rho}_A)+C_{l_1}(\rho_A))(C_{l_1}(\tilde{\rho}_A)-C_{l_1}(\rho_A))\\
&\leq& 2(N-1)(\sum^N_{\substack{
i,j=1\\
i\neq j}}\sqrt{p_ip_j}|\iinner{\phi_i}{\phi_j}|(1-|\iinner{u_i}{u_j}|))\\
&\leq& 2(N-1)(\sum^N_{\substack{
i,j=1\\
i\neq j}}\sqrt{p_ip_j}|\iinner{\phi_i}{\phi_j}|(\sqrt{1-|\iinner{u_i}{u_j}|^2}))\\
&\leq& 2(N-1)N(N-1)(\sum^N_{\substack{
i,j=1\\
i\neq j}}p_ip_j|\iinner{\phi_i}{\phi_j}|^2(1-|\iinner{u_i}{u_j}|^2)\\
&\leq& 2(N-1)^2N(\Tr{\rho^2_{AB}}-\Tr{\rho^2_A}),
\end{eqnarray*}
where the third line comes from the fact that $C_{l_1}(\tilde{\rho}_A), C_{l_1}(\rho_A)\leq N-1$.
Hence, the loss of coherence in  particle A depends on the entanglement between A and B. This
provides an interpretation for the decrease of duality bound in \eqref{eq:ECPdual}:
in the presence of quantum memory B, part of the coherence in
particle A is encoded in the entanglement between A and B, which leads to the loss of coherence
in the particle A and the decrease of the bound for duality relation between coherence and path information.

The duality relation \eqref{eq:ECPdual} also provides a tighter bound on duality relation \eqref{eq:CPdual}
for mixed states without quantum memory. Suppose a particle A  goes through an N-path interferometer  while
the initial state of particle A is a mixed state $\rho^0_A$. The orthonormal basis states $\set{\ket{i}}^N_{i=1}$ correspond
to the N possible  paths. Then there exists another quantum system B such that the bipartite state between A and B can be expressed as
in \eqref{eq:iniS}. Thus the initial density matrix of particle A is
\begin{eqnarray*}
\rho^0_A=\sum^N_{i,j=1}\sqrt{p_ip_j}\iinner{u_j}{u_i}\ket{i}\bra{j}_A.
\end{eqnarray*}
After the interaction with detector, the bipartite state between A and D is given by
\begin{eqnarray*}
\rho_{AD}&=&U\pa{\rho^0_A\ot \proj{\phi_0}_D}U^\dag\\
&=&\sum^{N}_{i,j=1}\sqrt{p_ip_j}\iinner{u_j}{u_i}\ket{i}\bra{j}_A\ot\ket{\phi_i}\bra{\phi_j}_D,
\end{eqnarray*}
where $U(\ket{i}_A\ket{\phi_0}_D)=\ket{i}_A\ket{\phi_i}_D$. Then the reduced density matrix of particle A and detector D are given by \eqref{den_A} and \eqref{den_D}, respectively.
Therefore, according to \eqref{ineq:immed1}, we get the duality relation for mixed state $\rho_A$ without quantum memory,
\begin{eqnarray}\label{eq:mixdual}
\nonumber&&\pa{P^A_s-\frac{1}{N}}^2+X^2_A\\
&\leq& \pa{1-\frac{1}{N}}^2
+2\frac{N-1}{N^2}\pa{\Tr{\rho^2_A}-\Tr{\rho^2_D}}.
\end{eqnarray}
Since $\Tr{\rho^2_A}\leq\Tr{\rho^2_D}$, the right-hand side of \eqref{eq:mixdual} is less than $\pa{1-1/N}^2$,
which provides a tighter bound than that of relation \eqref{eq:CPdual} for mixed states.
If $\rho^0_A$ is pure, $\Tr{\rho^2_A}$ and $\Tr{\rho^2_D}$ are equal to $\rho_A$ and $\rho_D$, which are the reduced
states of the pure state $\rho_{AD}=U\pa{\rho^0_A\ot \proj{\phi_0}_D}U^\dag$. Also, the relation \eqref{eq:mixdual}
becomes an equality for N=2.

Now, let us recall an entropic version of duality relation between path coherence
and the path information without quantum memory, that is, the duality relation between relative entropy coherence
of the particle A and the mutual information between detector states and measurement outcomes \cite{Bagan2016},
\begin{eqnarray}\label{ineq:En_dual}
I(D:M)+C_r(\rho_A) \leq H(\set{p_i}),
\end{eqnarray}
where $H(\set{p_i})=-\sum_i p_i\log p_i$ is the Shannon entropy, $C_r(\rho_A)$ is the relative entropy coherence of particle A, and $D, M$ are two random variables corresponding to the
detector states and the measurement outcomes of a POVM $\cM=\set{\Pi_i}^N_{i=1}$ respectively, where the joint distribution for $D,M$
is $p_{ij}=p(D=i,M=j)=p_i\Tr{\proj{\phi_i}\Pi_j}$ \cite{Bagan2016}. Note that the path information is quantified by
the mutual information $I(D:M)$ defined by
\begin{eqnarray}
I(D:M)=H(D)+H(M)-H(D,M),
\end{eqnarray}
where $H(D)=H(\set{p_i})$ and $H(M)=H(\set{q_j})$ with $q_j=\sum_ip_{ij}$.

In the following, we show an entropic duality relation between  coherence and path information in the presence of a quantum memory B,
\begin{eqnarray}\label{ineq:En_Edual}
I(D:M)+C_r(\rho_A)\leq H(\set{p_i})+S(B|A),
\end{eqnarray}
where $I(D:M)$ is the mutual information between detector states and measurement outcomes of a POVM $\cM=\set{\Pi_i}^N_{i=1}$ as defined in \cite{Bagan2016} and
the conditional entropy $S(B|A)=S(\rho_{AB})-S(\rho_A)$. The extra term $S(B|A)$ on the right-hand side of \eqref{ineq:En_Edual} quantifies the amount of entanglement between particle A and the memory B, as
$S(B|A)<0$ indicates the entanglement of $\rho_{AB}$ \cite{Cerf1997}.

Due to the presence of the quantum memory B, equation \eqref{den_w} is the state of the whole system after particle A interacts with the detector. Equations \eqref{den_A} and \eqref{den_D} are the reduced density matrix of particle $\rho_A$ and
$\rho_D$  respectively. The relative entropy of coherence for $\rho_A$ is given by
\begin{eqnarray}
C_r(\rho_A)=S(\rho^{\md}_A)-S(\rho_A)=H(\set{p_i})-S(\rho_A).
\end{eqnarray}
In view of the Holevo bound, the path information $I(D:M)$ is upper bounded as
\begin{eqnarray*}
I(D:M)\leq S(\rho_D)-\sum^N_{i=1}p_iS(\proj{\phi_i})
=S(\rho_D),
\end{eqnarray*}
where the von Neumann entropy for pure state is zero. Thus,
\begin{eqnarray}\label{ineq:immed2}
I(D:M)+C_r(\rho_A)
\leq H(\set{p_i})-S(\rho_A)+S(\rho_D).
\end{eqnarray}
Since $\rho_{AB}$ and $\rho_D$ are the reduced states of the pure state $\ket{\Psi}_{ABD}$,
the von Neumann entropy of $\rho_{AB}$ and $\rho_{D}$ are equal, $S(\rho_D)=S(\rho_{AB})$. Therefore, we obtain the duality relation  \eqref{ineq:En_Edual} for the case of the presence of a quantum memory. If $S(B|A)<0$, $\rho_{AB}$ is entangled and the right-hand side of \eqref{ineq:En_Edual} is strictly less than
$H(\set{p_i})$. That is, it provides a tighter bound than \eqref{ineq:En_dual} in this case. Also,
if the initial bipartite state $\ket{\psi}_{AB}$ between A and B is separable, $\ket{\psi}_{AB}=\ket{\psi'}_A\ket{u}_B$, where $\ket{u_i}$ in \eqref{eq:iniS} is equal to $\ket{u}$ up to a phase, then the
whole state of A,B and D after the interaction between particle A and D is of the form, $\ket{\Psi}_{ABD}=U(\ket{\psi'}_A\ket{\phi_0}_D)\ot \ket{u}_B$. Thus
$S(\rho_A)=S(\rho_{AB})$ and the duality relation \eqref{ineq:En_Edual} reduces to the relation \eqref{ineq:En_dual}.
Besides, as the accessible information is defined as $Acc(D)=\max_{POVM \cM}H(D:M)$ and the
duality relation holds for any POVM on detector state $\rho_D$, we obtain the following relation between
the accessible information and relative entropy of coherence in the presence of quantum memory,
\begin{eqnarray}
Acc(D)+C_r(\rho_A)\leq H(\set{p_i})+S(B|A).
\end{eqnarray}

For $N=2$, the von Neumman entropies of $\rho_A$ and $\rho_{AB}$ can be analytically calculated,
\begin{eqnarray*}
S(\rho_{AB})&=&-\pa{\frac{1}{2}+\lambda_1}\log\pa{\frac{1}{2}+\lambda_1}-\pa{\frac{1}{2}-\lambda_1}\log\pa{\frac{1}{2}-\lambda_1},\\
S(\rho_{A})&=&-\pa{\frac{1}{2}+\lambda_2}\log\pa{\frac{1}{2}+\lambda_2}-\pa{\frac{1}{2}-\lambda_2}\log\pa{\frac{1}{2}-\lambda_2},
\end{eqnarray*}
where
\begin{eqnarray*}
\lambda_1&=&\pa{\frac{p_1-p_2}{2}}^2+p_1p_2|\iinner{\phi_1}{\phi_2}|^2,\\
\lambda_2&=&\pa{\frac{p_1-p_2}{2}}^2+p_1p_2|\iinner{\phi_1}{\phi_2}|^2|\iinner{u_1}{u_2}|^2.
\end{eqnarray*}
As $\lambda_2\leq \lambda_1$, we have $S(\rho_A)\geq S(\rho_{AB})$ or $S(B|A)\leq 0$. Hence the right-hand side of \eqref{ineq:En_Edual} is less than
$H(\set{p_i})$ for two-path interferometer.

Similar to the case of $l_1$ norm measure, the duality relation \eqref{ineq:En_Edual} with quantum memory also gives a tighter bound
for mixed states without quantum memory. The duality relation for mixed states without quantum memory is described by the equation \eqref{ineq:immed2}, with the reduced density matrices of particle A and detector D given by \eqref{den_A} and \eqref{den_D}, respectively. It is easy to see that $S(\rho_A)\geq S(\rho_D)$ for $N=2$.

As another interesting scenario, we may also consider two entangled
particles A and B, such that A goes through an N-path interferometer and B goes through another.
Then coherence and path information of A and B
both satisfy the relation \eqref{eq:ECPdual}. We have
\begin{eqnarray}\label{ineq:sum}\nonumber
&&\pa{P^A_s-\frac{1}{N}}^2+\pa{P^B_s-\frac{1}{N}}^2+X^2_A+X^2_B\\\nonumber
&\leq& 2\pa{1-\frac{1}{N}}^2+\frac{2(N-1)}{N^2}\pa{\Tr{\rho^2_A}+\Tr{\rho^2_B}-2\Tr{\rho^2_{AB}}},
\end{eqnarray}
where $\rho_A$, $\rho_B$ and $\rho_{AB}$ are the reduced states of A, B and the system AB after the interaction
with the individual detectors. The relation \eqref{ineq:sum} becomes an inequality for $N=2$.

In conclusion, we have obtained two duality relations between path information and coherence for a particle
going through a multipath interferometer in the presence of a quantum memory, for both coherence quantifier
$l_1$ norm of coherence and relative entropy of coherence. We have shown that the entanglement between the particle and the quantum memory will lower down
the upper bounds of these duality relations, due to the decrease of coherence in the presence of entanglement.
Moreover, our bonds for wave-particle duality relations with quantum memory also provide the corresponding bonds for particles
in mixed initial states without quantum memory.
These results provide a new insight into the wave-particle duality and reveal
the role of quantum entanglement in the wave-particle duality.

\begin{acknowledgments}
K.F. Bu acknowledgements  Prof. Heng Fan, Dr. Yi Peng  and Yuming Guo for informative discussion on this topic during his visit in Beijing.
This work is supported by the Natural Science Foundation of China (Grants No. 11171301, No. 10771191, No. 11571307, and No. 11675113) and the Doctoral Programs Foundation of the Ministry of Education of China (Grant No. J20130061).
\end{acknowledgments}

\bibliographystyle{apsrev4-1}

 \bibliography{Dualcoh-lit}

\end{document}